\documentclass[twocolumn,floatfix]{revtex4-1}
\usepackage[utf8]{inputenc}
\usepackage{amsfonts,amsmath,amssymb,graphicx,epstopdf,verbatim}
\usepackage{placeins}
\usepackage{rotating}
\usepackage{epstopdf}
\usepackage{xcolor}

\begin{document}

\title{Cluster Gutzwiller Monte Carlo approach for a critical dissipative spin model}
\author{D. Huybrechts and M. Wouters}
\affiliation{TQC, Universiteit Antwerpen, Universiteitsplein 1, B-2610 Antwerpen, Belgium}

\begin{abstract}
    We study the influence of short-range quantum correlations and classical spatial correlations on the phase diagram of the dissipative XYZ model by using a Gutzwiller Monte carlo method and a cluster Gutzwiller ansatz for the wave function. Considering lattices of finite size we show the emergence of a ferromagnetic phase, two paramagnetic phases and the possible existence of a phase transition which is entirely quantum in nature. The inclusion of short-range quantum correlations has a drastic effect on the phase diagram but our results show the inclusion of long-range quantum correlations or the use of more sophisticated methods are needed to quantitatively match the exact results. A study of the susceptibility tensor shows that reciprocity is broken, a feature not observed in closed quantum systems. 
\end{abstract}

\date{\today}

\maketitle

\section{Introduction}

In recent years the field of dissipative phase transitions has been under intense study due to technological advances which have given access to a wide range of systems where it is possible to study the quantum dissipative behaviour of open quantum systems. The experimental platforms include trapped ions, cold atoms, semiconductor microcavities, cavity and circuit QED \cite{Bloch, Carusotto, Hartmann, Noh}. Driven-dissipative many-body systems find the origin of their interesting dynamics in the interplay of coherent driving and dissipation. 
Typically, in the long time limit, the system is driven into a steady state that is determined by the competition between both Hamiltonian and dissipative dynamics, resulting in a wide range of possible phases. Dissipative phase transitions have been observed in a wide range of experiments \cite{Rodriguez, Fitzpatrick, Baumann1, Baumann2, Brennecke, Fink1, Fink2}. Consequently, the interest in the study of dissipative phase transitions has spiked \cite{Casteels0, Hwang, Tomadin1, Leboite, Diehl1, Tomadin2, Biella, Rota, Overbeck2, Lee}. 

Analytical solutions for open quantum systems are scarce and if one wants to calculate the dynamics of the system, one has to rely on numerical tools. As usual, a numerically exact description is unfeasible due to the exponential scaling of the Hilbert space dimension with system size and approximations are needed.
Different approaches have been applied to a range of open quantum systems. 
Among them are the mean-field approximations \cite{Nissen, Jin1, Tomadin1, Owen, Leboite, Lee, Jin, Diehl1, Tomadin2, Biella}, the approaches based on matrix product operators and matrix product states \cite{Schroder, Manzoni, Orus, Verstraete1, Schollwoeck, Mascarenhas1, Cui, Verstraete2, Zwolak} - which have proven to be of great success in one dimension -, the corner space renormalization method \cite{Finazzi, Rota}, variational methods for the master equation \cite{Weimer, Overbeck1, McCutcheon, Pollock, Suri, Overbeck2} and variational methods at the level of the wave function \cite{Casteels1, Casteels2, Pichler, Verstraelen, Mascarenhas2, Nagy}. 

In this work we will study the dissipative XYZ Heisenberg model. Previous works have investigated the steady-state phase diagram by using a mean-field approach \cite{Lee}, a cluster mean-field approach (CMF) \cite{Jin} and the corner space renormalization method \cite{Rota}. The CMF has also been used together with a self-consistent Mori projector approach to investigate the (non)existence of limit cycles \cite{Owen}. A tensor network algorithm \cite{Kshetrimayum} and a driven-dissipative quantum Monte Carlo \cite{Nagy} have also been applied to this model. In Ref. \cite{Casteels2} this model and its phase transitions are studied using quantum trajectories and a single site Gutzwiller ansatz by using the Gutzwiller Monte Carlo approach (GMC). In this work we will extend this approach by applying a cluster Gutzwiller ansatz for the wave function. We will refer to this method as the cluster Gutzwiller Monte Carlo approach (CGMC).
By including quantum correlations we show the possible existence of a dissipative phase transition entirely quantum in nature, and not predicted by the classical mean-field method, through studying the steady-state spin structure factor and the susceptibility.

We will introduce the model we will be working with in section \ref{modelsystem}. The method of quantum trajectories, the used unravelling and the cluster Gutzwiller ansatz are explained in section \ref{qtraj}. In section \ref{Ss} we discuss the steady-state spin structure factor. The influence of an applied magnetic field will be studied in section \ref{appfield}. In section \ref{susc} the susceptibility and the possible existence of a new phase transition is looked into. Finally, conclusions are formulated in section \ref{conc}.

\section{The model system}\label{modelsystem}

We study the anisotropic XYZ Heisenberg Hamiltonian (with $\hbar=1$) on a spin lattice
\begin{equation}
\hat{H} = \sum_{\langle i,j\rangle}\left(J_{x}\hat{\sigma}_{i}^{(x)}\hat{\sigma}_{j}^{(x)}+J_{y}\hat{\sigma}_{i}^{(y)}\hat{\sigma}_{j}^{(y)}+J_{z}\hat{\sigma}_{i}^{(z)}\hat{\sigma}_{j}^{(z)}\right), \label{Hamiltoniaan}
\end{equation}
where $J_x$, $J_y$ and $J_z$ are the coupling strengths in the $x$, $y$ -and $z$-direction, $\hat{\sigma}_i^\alpha$ the Pauli matrices ($\alpha = x,y, z$) and the sum goes over the nearest neighbours in the lattice. Periodic boundary conditions are applied at the edge of the lattice. A proposal for an experimental setup based on the optical pumping of two-level atoms has been made in Ref. \cite{Lee} resulting in the above Hamiltonian with effective spins. This Hamiltonian governs the unitary part of the time evolution of the system. The total time evolution of the system is governed by a Lindblad Equation with dissipation along the z-axis
\begin{equation}
\partial_{t}\hat{\rho} = -i\left[\hat{H},\hat{\rho}\right]+\frac{\gamma}{2}\sum_{j}\left(2\hat{\sigma}_{j}^{(-)}\hat{\rho}\hat{\sigma}_{j}^{(+)}-\left\{\hat{\sigma}_{j}^{(+)}\hat{\sigma}_{j}^{(-)},\hat{\rho}\right\}\right).
\end{equation}
With $\gamma$ the decay rate of the spins and $\hat{\sigma}_i^{(+)}$ ($\hat{\sigma}_i^{(-)}$) the raising (lowering) operators along the z-axis. This driven-dissipative Heisenberg model has recently been subject of several studies attempting to describe the dynamics. In this work we will compare our results with the CMF \cite{Jin}. We will consider the parameters $J_x = 0.9\gamma$ and $J_z = \gamma$, unless stated differently, and vary $J_y$. The mean-field approach \cite{Lee} predicts a transition from a paramagnetic phase to a ferromagnetic phase for this parameter set at $J_y \approx 1.04\gamma$. By including quantum correlations in the CMF \cite{Jin} or classical spatial correlations in the GMC \cite{Casteels2} the existence of another transition from the ferromagnetic phase to the paramagnetic phase is observed.

\section{Quantum trajectories and the wave function ansatz}\label{qtraj}
The theory of quantum stochastic processes was first introduced by Davies \cite{Davies} and has been further developed \cite{Barchielli, Dalibard, Dum, Molmer, Carmichael1, Carmichael2} into the quantum trajectory formalism, also known as the Monte Carlo wave function method. It offers an alternative to calculating the dynamics of the system with a Lindblad equation for the density matrix of the open system. Instead of solving for the density matrix, the dynamics of the system are calculated on the level of the wave function. Through a stochastic process, many realizations of the wave function, so called quantum trajectories, are simulated and by averaging over these trajectories one recovers the dynamics of the open quantum system. This method is often referred to as the unravelling of the master equation. The stochastic process finds its origin in the continuous measurement of the environment. This continuous measurement results in random changes of the wave function of the system. The unravelling of the master equation is not unique, as there are several possible measurements of the environment. In this work we will unravel the master equation through the process of photon counting \cite{Breuer}. Usually the environment is under constant observation for emitted photons in this measurement scheme. As we are working with a spin system we will not be counting the excitations of the environment caused by emitted photons, we do however monitor the environment for excitations caused by spin-flips in the system. In between the detection of these excitations the wave function will evolve according to
\begin{equation}\label{psievolve}
    \psi(t) = \frac{\exp\left(-iHt\right)\Tilde{\psi}}{\vert\vert\exp\left(-iHt\right)\Tilde{\psi}\vert\vert},
\end{equation}
with $\Tilde{\psi}$ an initial (normalized) wave function. Note that the time evolution generated by 
\begin{equation}
H = \hat{H} - i\frac{\gamma}{2}\sum_i \hat{\sigma}_i^{(+)}\hat{\sigma}_i^{(-)},  
\end{equation}
does not preserve the norm. When an excitation is detected a quantum jump is made in the evolution of the wave function
\begin{equation}
    \psi \rightarrow \frac{\hat{\sigma}_i^{(-)}\psi}{\vert\vert\hat{\sigma}_i^{(-)}\psi\vert\vert},
\end{equation}
after which the wave function continues evolving according to (\ref{psievolve}).

Numerically solving the master equation for the solution of the density matrix is computationally very demanding. Exact solutions are unfeasible already for small systems. One of the reasons for this inconvenient characteristic of open many-body systems is the exponentially large Hilbert space. This is were one makes some efficiency gain by using quantum trajectories: in the master equation approach one has to work with the quadratically large Hilbert space $\mathcal{H}^2$ of the density matrix, whereas in the trajectory approach one works with the Hilbert space $\mathcal{H}$ of the wave function. This advantage comes at the cost of needing to average over multiple realizations to obtain the dynamics of the open system. The number of realizations needed, however, is usually much smaller than the dimension of the Hilbert space.
\begin{figure}
  \centering
    \includegraphics[width=0.33\textwidth]{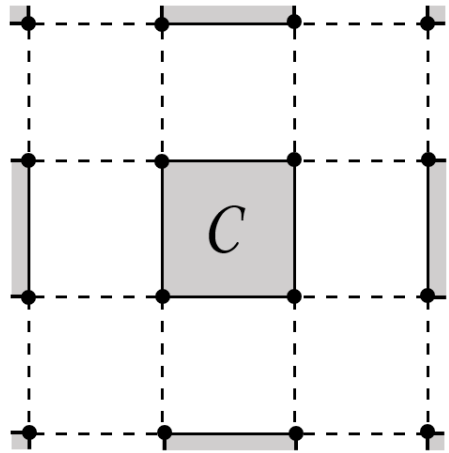}
      \caption{This figure shows the lattice layout with $2\times2$ clusters on a two dimensional lattice. Each cluster is shown as a grey area and contains a set of ('connected') lattice points. Inside these clusters quantum correlations between the different sites are included.}\label{clusterorientation}
\end{figure}

The quantum trajectory method however does not solve the exponential scaling of computational resources with system size. In order to reduce the dimensionality of the problem, an ansatz for the wave function will be considered. Previous work has investigated the Gutzwiller ansatz \cite{Gutzwiller} applied to the dissipative XYZ Heisenberg model \cite{Casteels2}. We extend this approach by including short-range quantum correlations through the use of the cluster Gutzwiller ansatz. This ansatz is realized by considering a sublattice of clusters, as shown in figure \ref{clusterorientation}, where each cluster contains a set of (nearest neighbour) lattice points
\begin{equation}
    \Psi_{GW}\left(\left\{\mathcal{C}\right\}\right) = \prod_{\mathcal{C}}\psi_{\mathcal{C}},
\end{equation}
where the product runs over the different clusters $\mathcal{C}$ with cluster wave function $\psi_\mathcal{C}$. This approach allows for the inclusion of both classical and quantum correlations inside the clusters and only classical correlations between the clusters. To keep the dimension of the Hilbert space limited only small clusters will be studied and the importance of short-range quantum correlations will be determined.

\section{Steady-state spin structure factor} \label{Ss}
In order to investigate the dissipative phase transition between a paramagnetic and ferromagnetic state, we will consider the steady-state spin structure factor $S_{SS}^{xx}(\textbf{k}=0)$, where:
\begin{equation}
S_{SS}^{xx}(\textbf{k}) = \frac{1}{N(N-1)}\sum_{j\neq l}e^{i\textbf{k}.(\textbf{j}-\textbf{l})}\langle\hat{\sigma}_{j}^{(x)}\hat{\sigma}_{l}^{(x)}\rangle.
\label{Sssxx}
\end{equation}
We use this correlation function rather than the spontaneous magnetization itself, because in a finite system, the $\mathbb{Z}_2$-symmetry does not spontaneously break. Alternatively, a (small) magnetic field could be applied to break the symmetry, as will be discussed in section \ref{appfield}. 

A non-zero value of the steady-state spin structure factor indicates a ferromagnetic phase. A zero value has a wider range of possibilities such as a paramagnetic phase, an anti-ferromagnetic phase and spin density waves. To distinguish between these phases different values of the wave vector have to be studied.

\begin{figure}
  \centering
    \includegraphics[width=0.5\textwidth]{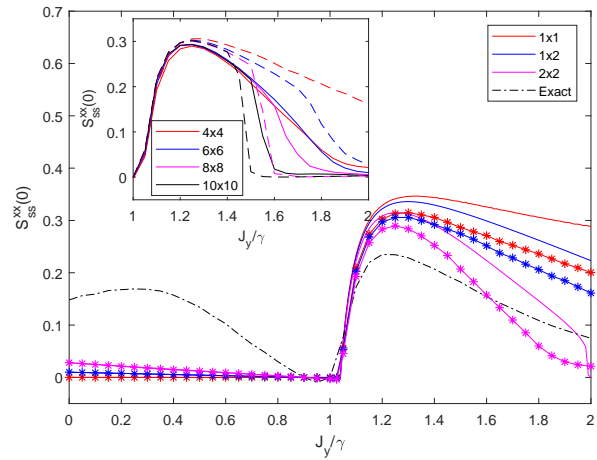}
      \caption{(Color online) Steady-state spin structure factor of a $4\times4$ lattice with different cluster sizes for the CGMC (stars) and the CMF (full lines). Note that the $1\times1$ CMF is the usual mean-field result. The CGMC qualitatively predicts the same ferromagnetic and two paramagnetic phases as the CMF \cite{Jin1}. Additionally we observe the possible existence of a phase transition completely quantum in nature for values of $J_y < 0.9\gamma$. \\ 
      Inset: steady-state spin structure factor of a $4\times4$, $6\times6$, $8\times8$ and $10\times10$ lattice with clusters of size $1\times2$ (dashed lines) and $2\times2$ (full lines) for the CGMC. Increasing the lattice size shows a sharpening of the phase transition, also found with the GMC \cite{Casteels2}. This sharpening is steeper when larger cluster sizes are included.}\label{Ssthermo}
\end{figure}

We simulate the dynamics of a trajectory over a minimum total time of $10.000/\gamma$ and obtain the steady-state solutions by time averaging over this single trajectory. 
The results of our numerical simulations are shown in Fig. \ref{Ssthermo}, where the spin structure factor was obtained for a 4x4 lattice with various cluster sizes. It is clear that increased incorporation of quantum correlations present for larger cluster sizes significantly affects the spin correlations.  

 Both the $1\times2$ and $2\times2$ clusters show the existence of the ferromagnetic phase and show qualitatively the same behaviour as predicted by the CMF \cite{Jin1} and the GMC \cite{Casteels2}. The clusters however, are able to find a non-zero value for $S_{ss}^{xx}(0)$ for values of $J_y < 0.9\gamma$. This behaviour is not captured by the single-site Gutzwiller ansatz \cite{Casteels2} or the mean-field \cite{Lee} and thus originates from quantum correlations. In the GMC \cite{Casteels2} the ferromagnetic region becomes smaller with growing system size and the transition to the paramagnetic phase becomes sharper. The inset of Fig. \ref{Ssthermo} shows that by including clusters of size $1\times2$ and $2\times2$ this behaviour is captured already for smaller lattice sizes. Increasing the cluster size makes the sharpening steeper and occur for smaller values of $J_y$. This shows the importance of quantum correlations in the simulation of an open quantum system. This sharpening is what is to be expected when the size of the lattice grows (i.e. when the thermodynamic limit is approached).

By comparing the results for different cluster sizes with the exact solution of this lattice we see that for increasing cluster sizes the exact solution is approached more closely, but differences persist. It has to be noted that short-range quantum correlations are not enough to accurately approximate the exact solution for the $4\times4$ system. It remains to be seen if this stays true for larger lattices and if longer-range quantum correlations have to be taken into account as well. 

As mentioned earlier, for values of the parameter $J_y$ smaller than $0.9\gamma$ we find an unexpected buildup of spin-spin correlations. Where the mean-field theory predicts an all-zero steady-state spin structure factor we find a non-zero value by including clusters. This non-zero value is most pronounced in the exact solution. This behaviour is completely neglected by the classical mean-field theory and thus entirely driven by quantum fluctuations.
The question remains whether a phase transition is present or not. It is however clear that short-range quantum correlations do not capture the exact behaviour for small lattices and long-range correlations have to be included for a more accurate description.

A comparison with the CMF used in Ref. \cite{Jin} can show us the importance of spatial correlations between the clusters, as they are not captured by the CMF. Fig. \ref{Ssthermo} shows the steady-state spin structure factor for several cluster sizes on a $4\times4$ lattice. We notice two distinct areas, again for $J_y < 0.9\gamma$ and $J_y > 0.9\gamma$ (we will resp. call them the \emph{left} hand and \emph{right} hand side). On the left hand side both the CMF and the CGMC match closely, giving a strong indication that only quantum correlations are important in this regime. On the right hand side of the figure however,  the CMF and CGMC match only qualitatively, in this regime both quantum and classical spatial correlations contribute. This confirms the difference between the buildup of order on the left and right hand sides respectively. \\ 
\begin{figure}
  \centering
    \includegraphics[width=0.5\textwidth]{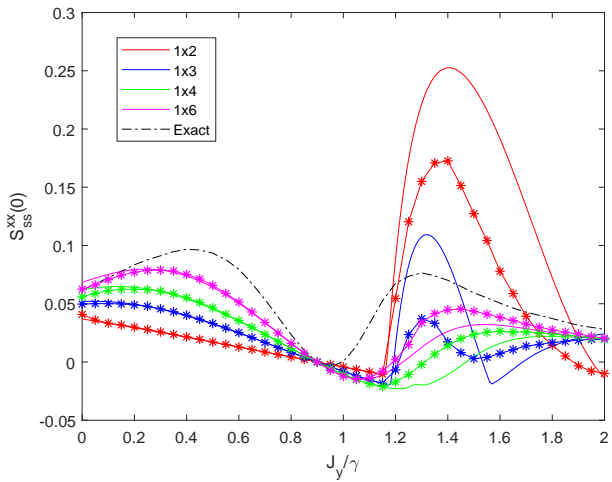}
      \caption{(Color online) Steady-state spin structure factor of a $1\times12$ lattice with different cluster sizes using the CGMC (stars) and the CMF (full lines). The figure shows that through inclusion of longer-range quantum correlations, by increasing the cluster size, the exact behaviour is approached more closely for $J_y < \gamma$. Increasing the cluster size for values of $J_y > \gamma$ does not show a clear convergence pattern to the exact solution. The results for $1\times6$ clusters do however match the exact result most closely. These findings, both for $J_y < \gamma$ and $J_y >\gamma$, indicate short-range quantum correlations are not sufficient for the description of the system.} \label{ss1x12}
\end{figure}

In 1D we can take clusters with larger linear size. In the top panel of Fig. \ref{ss1x12} we look at a $1\times12$ lattice for which we can go up to $1\times6$ clusters. We find that the behaviour of the steady-state spin structure factor is qualitatively captured by the cluster approach both in the left and right region for $1\times6$ clusters. It has to be noted that the system has no phase transition in the thermodynamic limit in 1D, which has been shown by using a matrix product operator ansatz for the density matrix \cite{Jin}. The behavior of the spin structure factor for finite size systems does however gives insight in the importance of the longer-range quantum correlations to describe the exact dynamics of the open quantum system. As one can see in Fig. \ref{ss1x12}, decreasing the size of the clusters results in a steady-state spin structure factor that differs completely from the exact value, even negative values are found for values of $J_y > \gamma$ where they should be positive. For values of $J_y < \gamma$ the influence of increasing the cluster size can be clearly observed. By including longer-range quantum correlations the exact behaviour is matched more closely. For $1\times2$ clusters we find the same linear behaviour for the steady-state spin structure factor as found in the 2D case. By increasing the cluster size we can see a clear convergence to the same behaviour as the exact solution. Short-range quantum correlations are as such not sufficient for the description of the system and longer-range quantum correlations play an important role.\\

\begin{figure}
    \centering
    \includegraphics[width = 0.5\textwidth]{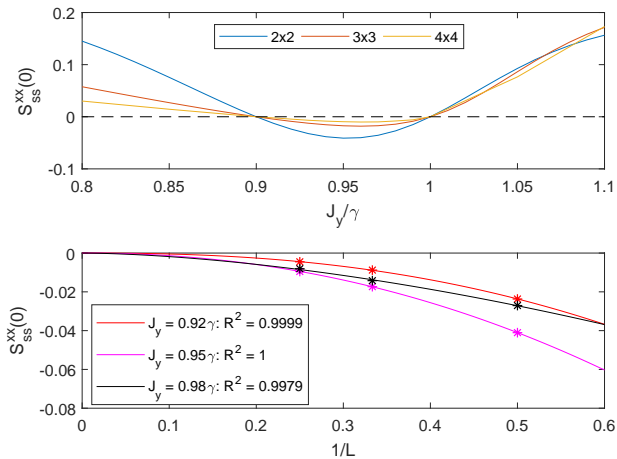}
    \caption{(Color online) Top panel: the exact solution for the steady-state spin structure factor of a $2\times2$, $3\times3$ and $4\times4$ lattice. To exclude the possibility of the presence of a continuous variation of the order parameter rather than a phase transition driven by quantum correlations, we study if the steady-state spin structure factor in the region $J_y \in ]J_x,J_z[ = ]0.9\gamma, \gamma[$ converges to zero in the thermodynamic limit. The solution of the $2\times2$ and $3\times3$ was found by solving the master equation and the $4\times4$ solution was obtained with the trajectory approach.\\
    Bottom panel: The behaviour of the steady-state spin structure factor through a fit as a function of lattice size from the known points of the $2\times2$, $3\times3$ and $4\times4$ lattice for several values of $J_y \in ]J_x,J_z[$. The fit of a power law of the form $S_{ss}^{xx}(0, L) = aL^b$ for $L\times L$ lattices returns high $R^2$-values and converges to zero in the thermodynamic limit.}
    \label{fig:exactsol}
\end{figure}

To further confirm the existence of the phase transition driven by quantum correlations we show the steady-state spin structure factor for the exact solution of a $2\times2$, $3\times3$ and $4\times4$ lattice on the top panel of figure \ref{fig:exactsol}. As can be seen from this figure, and as is expected for finite size systems, the region where the phase transition occurs is smoothed out and one could suspect that there is only a continuous change of the order parameter rather than an actual phase transition. In the parameter region $J_y/\gamma \in ]0.9,1[$ however the spin structure factor does decrease when the lattice size is increased ( Note that the steady-state spin structure factor is always zero for the values $J_y = J_x$ and $J_y = J_z$ due to the unitary dynamics conserving respectively the magnetization in the z-direction and x-direction \cite{Jin}). In order to check for the convergence as a function of increasing system size, we show in the bottom panel of Fig. \ref{fig:exactsol} the behavior as a function of the system size, together with a fit to the power law dependence $S_{ss}^{xx}(0, L) = aL^b$, where $L$ is the number of points along one dimension of the $L\times L$ lattice. In all cases, we find a negative exponent $b$, which is compatible with a vanishing of the spin structure factor in the thermodynamic limit. For $J_y=0.95\gamma$ (in the middle of the interval), we find $b\approx -2.1$, close to the value $b=-2$, that is expected for a two-dimensional system with a finite correlation length.

\section{An applied magnetic field}\label{appfield}
The $\mathbb{Z}_2$-symmetry can be explicitly broken by applying a small magnetic field. 
In this section we will study the behaviour of the magnetization of the system as a function of the applied field in the $x$-direction and $y$-direction. An applied field $\vec{h} = h_x\vec{e}_x+h_y\vec{e}_y$ translates in adding a term $\hat{H}_B$ to the Hamiltonian $\hat{H}$ from (\ref{Hamiltoniaan})
\begin{equation}\label{Hfield}
    \hat{H}_B = h\sum_i\left(\cos(\theta)\hat{\sigma}_i^{(x)}+\sin(\theta)\hat{\sigma}_i^{(y)}\right).
\end{equation}

\begin{figure}
  \centering
    \includegraphics[width=0.5\textwidth]{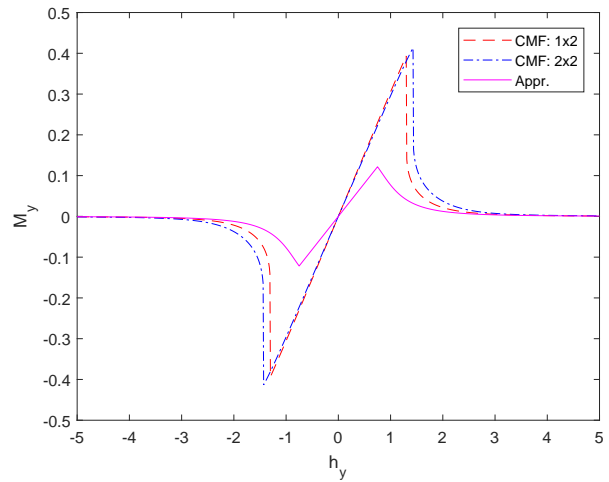}
      \caption{(Color online) Magnetization in the y-direction as function of an applied magnetic field in the y-direction for $1\times2$ (dashed red line) and $2\times2$ (dash-dotted blue line) clusters in the CMF and the approximated mean-field solution (full purple line) which is valid in the limit of large $h_y$ for $J_y=0.25\gamma$.} \label{AppField}
\end{figure}
In Fig. \ref{AppField} the magnetization in the $y$-direction is shown for a $4\times4$ lattice with $1\times2$ and $2\times2$ clusters as a function of $h_y$ $(h_x=0)$ in the CMF. 
From the theory of closed quantum systems one would expect the magnetization to saturate at $\pm 1$. This however is not the case as can be seen in the figure, both the $x$ and $y$ magnetization go to zero when the field is increased. To obtain a closer understanding of this behaviour we note that it is also present in the mean-field approximation, for which we can find analytic expressions. These expressions will enable us to shed light on this behaviour.\\
The system of mean-field equations in the steady-state can be written as
\begin{equation}
\left\{
\begin{split}
&-\frac{\gamma}{2}M_x+2d(J_y-J_z)M_yM_z+2h_yM_z = 0, \\
&-\frac{\gamma}{2}M_y+2d(J_z-J_x)M_xM_z-2h_xM_z = 0, \\
&-\gamma\left(M_z+1\right)+2d(J_x-J_y)M_xM_y\\
&+2(h_xM_y-h_yM_x) = 0,
\end{split}\label{mfeq}
\right.
\end{equation}
with $d$ the dimensionality, $\gamma$ the dissipation rate and $M_x$, $M_y$ and $M_z$ resp. the magnetization in the $x$, $y$ and $z$ direction. We look at the case where $h_y \neq 0$ and $h_x = 0$ (the reverse situation is analogue). With these parameters one can rewrite the system of equations as an expression for $M_x$ and $M_y$ in terms of $M_z$ 
\begin{equation}
M_x = \frac{1}{\gamma}\frac{4h_y M_z}{1 - \frac{16d^2}{\gamma^2}\left(J_y-J_z\right)\left(J_z-J_x\right)M_z^2},
\end{equation}
\begin{equation}
M_y = \frac{4d}{\gamma}(J_z-J_x)M_xM_z,
\end{equation}
and an equation for $M_z$ of which the solution remains to be found by substituting the above solutions for $M_x$ and $M_y$ into the last equation in \eqref{mfeq}. This equation has no analytic solution and has to be solved numerically. We can however use the knowledge that $M_z \rightarrow 0$ as $h_y \rightarrow \infty$. A more close study shows that for a growing field $h_y$, $h_yM_z  \rightarrow 0$. With these conditions we can approximate the third equation of \eqref{mfeq} up till order $M_z$. We then find for large $h_y$
\begin{equation}
M_z = -\frac{1}{8}\frac{\gamma^2}{h_y^2}.
\end{equation}
This relation explains why $M_x$ and $M_y$ go to zero for big applied fields, rather than $\pm 1$. This is shown in Fig. \ref{AppField} as the magenta line.
Unlike in thermal equilibrium, the magnetization goes to zero for large magnetic field. The reason is that in the limit $\vert h_y\vert \rightarrow \infty$, the Zeeman term dominates the Hamiltonian, so that the eigenstates are simply the eigenstates of $\sigma_y$. The dissipation being in the orthogonal direction does not drive the system to the ground state, but rather destroys the coherence between the eigenstates. The resulting steady state is then the unit matrix, from which the zero magnetization follows.

For the special case of $J_x=J_y=J_z$ ($\gamma = 1$) the system of mean-field equations is analytically solvable without any approximations:
\begin{equation}
\left\{
\begin{split}
&M_x = - \frac{h_y}{\frac{1}{4} + 2\left(h_x^2+h_y^2\right)}, \\
&M_y = \frac{h_x}{\frac{1}{4} + 2\left(h_x^2+h_y^2\right)}, \\
&M_z = - \frac{1}{1 + 8 \left(h_x^2+h_y^2\right)}.
\end{split}
\right.
\label{eq:Mmf}
\end{equation}
This is in agreement with the above result and is also true for small $h_x$ and $h_y$. We can conclude that a large magnetic field will cause the system to have no magnetization at all.

Note the difference in sign between $M_x$ and $M_y$ in this special case. This sign difference implicates that we cannot interchange $x$ and $y$ without introducing a sign change. At first sight this might appear confusing because of the identical parameters $J_x = J_y = J_z$. Only the $z$-direction is fixed by the dissipation and so there appears to be no clear reason for a distinction between $x$ and $y$. A closer look shows that one cannot interchange $x$ and $y$ because this changes the handedness of our coordinate system. This is reflected in the commutation relations of the Pauli matrices requiring that  $\left[\hat{\sigma}^{(a)},\hat{\sigma}^{(b)}\right] = 2i\epsilon_{abc}\hat{\sigma}^{(c)}$. To interchange $x\rightarrow y$ one could do $\hat{\sigma}^{(x)} \rightarrow \hat{\sigma}^{(x)}$ and $\hat{\sigma}^{(y)} \rightarrow -\hat{\sigma}^{(y)}$. This would however result in $\hat{\sigma}^{(z)} \rightarrow - \hat{\sigma}^{(z)}$ in order to satisfy the Pauli commutation relations. So $x$ and $y$ cannot be interchanged without changing the sign of $z$. This also results in the same coordinate system and thus no interchange was made in the end. As such there is no symmetry to transform $x$ into $y$ explaining why a sign difference can be present.
\section{Susceptibility}\label{susc}
\begin{figure}
  \centering
    \includegraphics[width=0.5\textwidth]{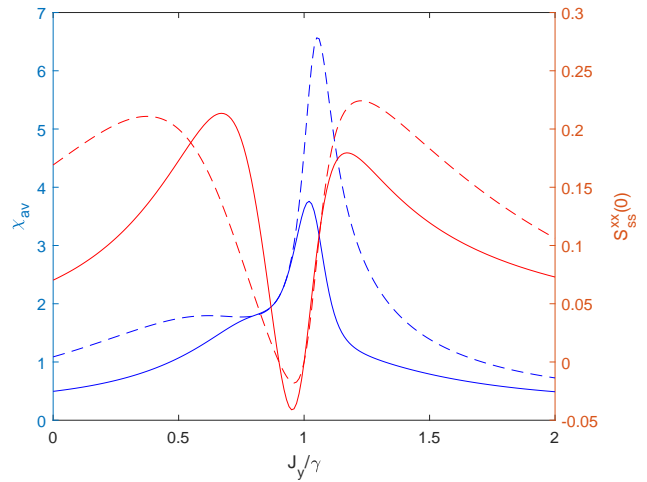}
      \caption{(Color online) Exact solution of the average angular susceptibility (blue) and steady-state spin structure factor (red) of a 2x2 (full line) and 3x3 (dashed line) lattice as a function of $J_y$. Both solutions show a 'shoulder' for $J_y < 0.9\gamma$. This shoulder could be a second peak in the susceptibility, masked by the higher peak on the right, suggesting a phase transition. The position of this peak corresponds to the non-zero region in the steady-state spin structure factor.} \label{Susc}
\end{figure}
A divergence in the susceptibility indicates the existence of a phase transition. To calculate the susceptibility we use the scheme presented in \cite{Rota} and apply a field in the $xy$-plane which corresponds to adding a term \eqref{Hfield} to the system Hamiltonian. The resulting magnetization is given by
\begin{equation}
\vec{M}(h,\theta) = 
\begin{pmatrix}
    \chi_{xx}       & \chi_{xy} \\
    \chi_{yx}       & \chi_{yy} 
\end{pmatrix}
.
\begin{pmatrix}
h\cos(\theta)\\
h\sin(\theta)
\end{pmatrix}.
\end{equation}
The susceptibility tensor can be extracted through
\begin{equation}
\chi_{\alpha\beta} = \left.\frac{\partial M_\alpha}{\partial h_\beta}\right\vert_{h= 0}.
\end{equation}
With $M_\alpha$ the magnetization in the $x$ or $y$ direction. To obtain a scalar value the average angular susceptibility can be calculated as follows
\begin{equation}
\chi_{av} = \frac{1}{2\pi}\int_0^{2\pi} d\theta \left.\frac{\partial\vert\vec{M}(h,\theta)\vert}{\partial h}\right\vert_{h=0}.
\end{equation}
For a more elaborate discussion we refer to \cite{Rota}.\\
When we calculate the susceptibility tensor for different cluster and lattice sizes we find that $\chi_{xy}\neq\chi_{yx}$. This is strikingly different from the case of closed systems, where the susceptibility is found from the free energy $F$, $\chi_{xy} = \frac{\partial^2 F}{\partial h_x\partial h_y} = \frac{\partial^2 F}{\partial h_y\partial h_x} = \chi_{yx}$.

We find this result even in the mean-field approximation. If we take the parameter values $J_x=J_y=J_z$ we find that $\chi_{xy} = - \chi_{yx}$  from \eqref{eq:Mmf}. 
Numerical results show that for general coupling parameters the magnitudes differ and in general we find  $\vert\chi_{xy}\vert\neq\vert\chi_{yx}\vert$ and reciprocity is broken.\\
For the 2D lattices we have two regions of interest. The earlier mentioned \emph{right} hand region, studied in \cite{Casteels2, Lee, Jin}, and secondly the \emph{left} hand region, discussed in section \ref{Ss}.
In Fig. \ref{Susc} the exact solution of the susceptibility for a $2\times2$ and $3\times3$ lattice is shown. Larger lattices are computationally not feasible and a more sophisticated method would have to be used, such as the Corner Space method \cite{Rota}. The right hand region of the susceptibility has been studied in \cite{Rota}. The presence of a peak in the susceptibility could indicate a phase transition. Note the `shoulder' which is present on the left side. This could indicate the presence of a second peak, partially masked by the higher peak on the right hand side. These two peaks move away from each other when the lattice size is increased. The peak on the left however is not sharp and it remains to be seen whether it diverges for larger lattice sizes and whether a true phase transition is present. It should be noted that the positions of the shoulder do coincide with a sharp decrease of the steady-state spin structure factor. These findings show the possible presence of a phase transition which is entirely quantum in nature, which is completely missed when the quantum correlations are neglected.

\section{Conclusions}\label{conc}
We studied the dissipative XYZ Heisenberg model with the cluster Gutzwiller Monte Carlo. This method allows for the inclusion of short-range quantum correlations as well as classical spatial correlations. Calculation of the steady-state spin structure factor shows the appearance of a ferromagnetic phase and two paramagnetic phases also found in Ref. \cite{Jin1} and Ref. \cite{Casteels2}. We show the possible existence of another phase transition which is entirely quantum in nature and which is completely missed when quantum correlations are neglected. A calculation of the susceptibility tensor shows how reciprocity is broken, a feature not observed in closed quantum systems. We show that the inclusion of short-range quantum correlations causes the sharpening of the phase transition to occur for smaller lattice sizes. Even though short-range quantum correlations have a big impact on the phase diagram of the system it has to be noted that they only qualitatively match the exact results and long-range quantum correlations play an important role in the dynamics of the system.

\acknowledgements  Discussions with C. Ciuti, F. Storme, W. Verstraelen and M. Van Regemortel are greatfully acknowledged. This work is supported by UAntwerpen/DOCPRO/34878. Part of the computational resources and services used in this work were provided by the VSC (Flemish Supercomputer Center), funded by the Research Foundation - Flanders (FWO) and the Flemish Government department EWI.

\end{document}